\documentclass[sensors,article,accept,moreauthors,pdftex]{Definitions/mdpi}

\usepackage{graphicx}
\usepackage{subcaption}
\usepackage[ruled, linesnumbered]{algorithm2e}
\firstpage{1} 
\pubvolume{xx}
\issuenum{1}
\articlenumber{5}
\pubyear{2019}
\copyrightyear{2019}
\history{Received: 15 December 2019; Accepted: 23 January 2020; Published: date}
\Title{Multi-Cell LTE-U/Wi-Fi Coexistence Evaluation Using a Reinforcement Learning Framework}

\Author{\hl{Jos}{\'e} M.~de C.~Neto $^{*}$\orcidA{},  %Please carefully check the accuracy of names and affiliations. Changes will not be possible after proofreading. 
Sildolfo F.~G.~Neto \orcidB{}, Pedro M. de Santana $^{\dagger}$\orcidC{} and Vicente~A.~\hl{de}~Sousa~Jr~\orcidD{}}
\AuthorNames{José Martins, Sildolfo Neto, Pedro Santana and Vicente Sousa}

\address{%
$^{1}$ \quad Federal University of Rio Grande do Norte, \hl{Natal-RN}, 59078-970, Brazil;  %please provide the post code
sildolfo@ufrn.edu.br (S.F.G.N.);
pedromaia02@ufrn.edu.br (P.M.d.S.);
vicente.sousa@ufrn.edu.br (V.A.D.S.J.)
}

\corres{Correspondence: martinsee@ufrn.edu.br%; Tel.: (optional; include country code; if there are multiple corresponding authors, add author initials) +xx-xxxx-xxx-xxxx (F.L.)
}

\firstnote{Current address: Robert Bosch GmbH Corporate Research, \hl{Renningen}, 71272,  %please provide the post code
Germany.} 
\abstract{Cellular broadband Internet of Things (IoT) applications are expected to keep growing year-by-year, generating demands from high throughput services. Since some of these applications are deployed over licensed mobile networks, as long term evolution (LTE), one already common problem is faced: the scarcity of licensed spectrum to cope with the increasing demand for data rate. The LTE-Unlicensed (LTE-U) forum, aiming to tackle this problem, proposed LTE-U to operate in the 5~GHz unlicensed spectrum. However, Wi-Fi is already the consolidated technology operating in this portion of the spectrum, besides the fact that new technologies for unlicensed band need mechanisms to promote fair coexistence with the legacy ones. In this work, we extend the literature by analyzing a multi-cell LTE-U/Wi-Fi coexistence scenario, with a high interference profile and data rates targeting a cellular broadband IoT deployment. Then, we propose a centralized, coordinated reinforcement learning framework to improve LTE-U/Wi-Fi aggregate data rates. The added value of the proposed solution is assessed by a ns-3 simulator, showing improvements not only in the overall system data rate but also in average user data rate, even with the high interference of a multi-cell environment.
}


\begin{document}
%%%%%%%%%%%%%%%%%%%%%%%%%%%%%%%%%%%%%%%%%%

%%%%%%%%%%%%%%%%%%%%%%%%%%%%%%%%%%%%%%%%%%
% \setcounter{section}{-1} %% Remove this when starting to work on the template.
% \section{How to Use this Template}
% The template details the sections that can be used in a manuscript. Note that the order and names of article sections may differ from the requirements of the journal (e.g., the positioning of the Materials and Methods section). Please check the instructions for authors page of the journal to verify the correct order and names. For any questions, please contact the editorial office of the journal or support@mdpi.com. For LaTeX related questions please contact latex@mdpi.com.
%The order of the section titles is: Introduction, Materials and Methods, Results, Discussion, Conclusions for these journals: aerospace,algorithms,antibodies,antioxidants,atmosphere,axioms,biomedicines,carbon,crystals,designs,diagnostics,environments,fermentation,fluids,forests,fractalfract,informatics,information,inventions,jfmk,jrfm,lubricants,neonatalscreening,neuroglia,particles,pharmaceutics,polymers,processes,technologies,viruses,vision

\section{Introduction}
% The introduction should briefly place the study in a broad context and highlight why it is important. It should define the purpose of the work and its significance. The current state of the research field should be reviewed carefully and key publications cited. Please highlight controversial and diverging hypotheses when necessary. Finally, briefly mention the main aim of the work and highlight the principal conclusions. As far as possible, please keep the introduction comprehensible to scientists outside your particular field of research. Citing a journal paper \cite{ref-journal}. And now citing a book reference \cite{ref-book}. Please use the command \citep{ref-journal} for the following MDPI journals, which use author-date citation: Administrative Sciences, Arts, Econometrics, Economies, Genealogy, Humanities, IJFS, JRFM, Languages, Laws, Religions, Risks, Social Sciences.

The increasing use of the wireless licensed spectrum, as well as its scarcity, comes with the massive demand from mobile data, as the number of wireless devices connected to the internet is expected to keep growing every day. According to~\cite{ericsson_report}, the number of mobile broadband subscriptions in the first quarter of 2019 reached 6 billion, corresponding to a growth rate of 15 percent year-on-year. In the cellular Internet of Things (IoT) applications (services using a cellular infrastructure with no need for a new private network), the same report also estimates that 4.4 billion devices will be anchored in broadband by the end of 2024.  These services that correspond to IoT deployments require a large amount of data, e.g., multimedia data, augmented reality (AR), and virtual reality (VR). 

Historically, previous cellular broadband IoT applications have been enabled by the third-generation (3G) of mobile communication. Still, with the increasing demand, the fourth generation (4G) rises as the primary enabler, since it can better cope with such demand. However, the spectrum of long term evolution (LTE) 4G deployments, or the LTE-M (modified LTE for IoT applications introduced in 3rd generation partnership project (3GPP) Release 13~\cite{3GPP_TR36_300}) needs licenses earned in an auction by the operators, which is expensive. Additionally, most of the licensed bands already have  an overused profile, which is impracticable for IoT deployment, especially for broadband services and a significant number of devices. Then, one alternative to overcome those drawbacks is to leverage the unlicensed spectrum once it is free to use, and to offer cheap additional bandwidth.

Nevertheless, some problems can be solved with this solution. Some regional regulatory entities in the world, like in Europe and Japan, require that devices using the unlicensed spectrum implement a Listen-Before-Talk (LBT) mechanism, while others do not, like in South Korea and the United States. Another problem is that Wi-Fi (IEEE 802.11 a/n/ac) is the main, and most successful, access technology already using the unlicensed spectrum. Wi-Fi is one of the first choices for some IoT deployments, besides the other non-cellular solutions like LoRaWAN~\cite{lora} and SigFox~\cite{sfox}. Thus, new technologies trying to exploit the unlicensed band must, somehow, take into account the better coexistence among the technologies so that there is no monopoly from one access technology. Three leading solutions were proposed to deal with these two problems: The LTE-Unlicensed (LTE-U), proposed by the LTE-U Forum~\cite{lte_u_forum}; LTE-Licensed Assisted Access (LAA) proposed by 3GPP release 13~\cite{3GPP_TR36_300}; MuLTEfire proposed by Qualcomm \cite{multefire}. 

LTE-U is \hl{a} LTE solution for the unlicensed band proposed by the LTE-U Forum (Qualcomm, Verizon, Samsung and Ericsson) targeted to work primarily in the 5 GHz ISM band in countries where regulatory entities do not require an LBT mechanism~\cite{lte_u_forum}. 3GPP 10 introduced carrier aggregation, which is the primary mechanism to enable this LTE solution in the unlicensed band. To equitably coexist with Wi-Fi, the LTE-U also implements a MAC layer mechanism based on a time division multiplexing (TDM)-like duty cycle (DC) approach. Targeting broadband IoT deployments, this duty cycle approach leads to a behavior where the energy consumption can even be reduced since the device with this technology operates following ON/OFF periods. But even with this feature, LTE-U is rarely explored, because of the Wi-Fi monopoly and the LTE-M solution for licensed spectrum.   

Targeting deployments in Europe and Japan, where an LBT mechanism is mandatory, 3GPP in its Release 13 proposed the LTE-Licensed Assisted Access (LAA)~\cite{TR36_889}. Like LTE-U, LTE-LAA uses a standard LTE with modifications to operate in the unlicensed band but with an anchor LTE carrier in the licensed spectrum. LAA implements an LBT mechanism very similar to the carrier-sense multiple access with collision avoidance (CSMA/CA) of the Wi-Fi, but with different parameters. On the other hand, MuLTEfire~\cite{multefire} is another proprietary solution proposed by Qualcomm aiming a stand-alone operation in the unlicensed spectrum, i.e., without a licensed anchor. However, compared to LTE-LAA, MulteFire Release 1.1 Specification has an expanded support for IoT services with Low power wide area at 800/900 MHz with NB-IoT-U (Narrowband IoT-Unlicensed), and 2.4 GHz with eMTC-U (enhanced machine type communication-unlicensed)~\cite{multiFSpecSite}.

As a branch of the 802 families by IEEE, the IEEE 802.11 is a standard targeted to wireless area local networks (WLANs) operating in the unlicensed band, mostly with 2.4~GHz and 5~GHz frequency~\cite{book_wifi}. Wi-Fi is the global brand name used by all the products based on IEEE 802.11 and certified by the Wi-Fi Alliance regarding interoperability as the standard evolves. As the demand and deployments of Wi-Fi grew up, new generations were standardized, bringing the creation of Wi-Fi 2, 3, 4 and 5,  based on the 802.11a, 802.11g, 802.11n and 802.11ac, respectively, to cope with even higher data rates for even higher demands. All these subsequent versions, except for Wi-Fi 3, use the 5~GHz band and adopt the orthogonal frequency division multiplexing (OFDM) technology, as LTE-LAA and LTE-U. But despite the similarities, the medium access control (MAC) layer of Wi-Fi differs from the ones in LTE-U/LAA, creating the necessity for investigations in the coexistence of such technologies, especially when targeting scenarios for broadband internet applications. 

This paper extends our previous coexistence solution~\cite{Dm_csat} to a multi-cell scenario with co-channel users. The herein proposed coexistence framework uses Q-Learning applied to LTE-U to automatically adapt the duty cycle parameter based on the transmitted data rate as the central metric for assessing coexistence interference. Our solution is a centralized, coordinated reinforcement learning framework to improve the system aggregate data rate.

This paper is organized as follows. Section~\ref{rel_works} presents a literature review on LTE-U/Wi-Fi coexistence with a focus on the application of machine learning techniques in the coexistence problem. We also highlight our main contributions comparing with those previous works. In Section~\ref{review}, we review the LTE-U and Wi-Fi as well as their main coexistence mechanisms. System modeling, scenario definition and preliminaries results are presented in Section~\ref{model}. The proposed framework is presented in Section~\ref{frame}. In Section~\ref{eval}, we present and discuss the results of our solution. Section~\ref{conc} presents the conclusions and the future perspectives of this work.  

\section{Related Works}
\label{rel_works}

We can find several works dealing with the coexistence of the new unlicensed LTE technologies and the consolidated Wi-Fi~\cite{p_eval_indians,laa_pedro,p_eval_globecom,laa_survey}. One common finding of these works is that the LTE solutions are, for the evaluated scenarios, a better neighbor for Wi-Fi than Wi-Fi itself in terms of interference and capacity. Another important finding states that this coexistence problem has desired characteristics for applying machine learning techniques to reduce the coexistence impact. In~\cite{qualcomm_tech}, the specific coexistence of LTE-U and Wi-Fi is analyzed. The main analysis of this work is how the adjustment of the duty cycle parameter influences the system throughput. 

Medium access mechanisms are key blocks to allow coexistence between different technologies in the unlicensed spectrum. Thus, the mismatch between parameters can lead one technology to overlap others concerning access to the communication channel~\cite{Zinno2018}. Even with this concern, LTE-LAA, with the carrier-sense multiple access with collision avoidance (CSMA-CA) mechanism, can achieve a performance equal to or better than Wi-Fi coexisting with Wi-Fi~\cite{laa_pedro}. With LTE-U, channel access alternatives based on channel sensing and dynamic adaptation of the duty cycle fractional work cycle have been proposed~\cite{fuad_coex18, paper_erika_2013}. The authors in~\cite{paper_erika_2013} proposed an approach to disable standard LTE transmissions on specific subframes using the almost blank subframe (ABS) LTE functionality. Their results demonstrate that LTE-DC can operate harmoniously with Wi-Fi and even LTE-LAA by adequately adjusting the duty cycle~\cite{Bojovic2019,Alhulayil2018,Santana2017}.

Some classic optimization techniques are used to enhance coexistence metrics. Attempting a better Quality of Service (QoS) level, a solution based on joint optimization of transmission time, carrier allocation and power is proposed ~\cite{Tan2019}. The authors also used throughput and delay-based user association as the metric for LAA/Wi-Fi coexistence. They achieved the theoretical performance with the solution, although there is trade-off between the desired QoS level and solution complexity. In~addition to QoS, fairness is another metric exploited by classical optimization techniques. The first algorithm proposed in~\cite{Maule2018}, based on connection admission control (CAC), is responsible for fairness in the sharing of resources. The second algorithm uses a duty cycle approach, similar to the one proposed for LTE-U, but applied to LTE-LAA to improve delay related metrics. Still in the LAA/Wi-Fi coexistence, the authors of~\cite{Mehrnoush2018} provide an analytical study of the influence of energy detection (ED) on the throughput of each technology. Conclusions indicate that maximum system throughput can be achieved by adjusting the ED threshold. A branch of classical approaches called game theory is also used to solve the problem of LTE/Wi-Fi coexistence. A coalition game-based framework, in a real deployment of commercial routers are used as a testbed of an LTE-U/Wi-Fi scenario. In this scenario, and with the proposed solution, the goal of users in the game is to maximize system throughput (sum of each user)~\cite{Zhang2018}. Another game, which is cooperative and based on the Nash bargain, is presented in~\cite{Bairagi2018}. LTE-U users QoS is enhanced by sharing resources between LTE-U and Wi-Fi while protecting Wi-Fi.

Coexistence solutions can be classified regarding the radio resource management (RRM) strategy to be modified to achieve the coexistence. We summarize the following list of solutions:

\begin{itemize}[leftmargin=*,labelsep=5.8mm]
    \item \textbf{Carrier detection and sensing:} The authors of~\cite{Ratasuk2012} implement adjustments on the LTE MAC layer to support this technique. After evaluations, they concluded that LTE presents throughput gains without decreasing the performance of Wi-Fi systems. Aiming channel access and QoS fairness, Li~et~al.~\cite{Li2015}~devised an enhanced LBT scheme to adaptively tune the clear channel assessment (CCA) threshold of LTE-LAA and interference avoidance to Wi-Fi. For LTE-U coexistence, two-channel sensing schemes are proposed in~\cite{Jia2015}, including periodic sensing and persistent sensing. In the periodic scheme, the LTE-U APs sense the medium for a fraction of time within each subframe and then determine if it is available to transmit data during the remaining subframe. In the persistent sensing scheme, the LTE-U APs also sense the channel during one entire subframe until the channel becomes free, and then transmits the data during the following several subframes. 
    \item \textbf{Link-adaptation:} Using the already defined Wi-Fi features to support coexistence, authors in~\cite{Bhorkar2015} proposed a link-adaptation algorithm based on Wi-Fi's MAC distributed coordination function (DCF) protocol to enhance LTE-LAA's throughput performance. The authors in~\cite{Jeon2014} use a scheme similar to Wi-Fi's request to send (RTS)/clear to send (CST) enabling protections for the LTE cells and helping Wi-Fi stations to save energy.  %Please define CST if appropriate. ok, added.
    \item \textbf{Power control:} In~\cite{Study2015}, the authors affirmed that the static control power used for Wi-Fi might be used in a similar form for LTE in unlicensed bands, controlling coexistence interference. The authors of~\cite{Chaves2013} compared the duty cycle mechanism with the power control mechanism for the LTE-U uplink, which resulted in a  higher average user throughput for both LTE-U and Wi-Fi compared to LTE-U with the specific duty cycle of 80\%. Regarding the ABS mechanism, they also affirm that this is a conservative solution, and the coexistence may be better addressed using some power control approach.
    \item \textbf{Spectrum slicing:}  The authors of~\cite{Teng2017} adopt the spectrum slicing, a dynamic and statics spectrum sharing through spectrum division into several partitions. In their solution, each partition can be exclusively accessed by one mobile network operator with few concerns about power control and therefore enhancing the coexistence scenario. Furthermore, works like~\cite{Xu2015} propose the use of game theory frameworks to mitigate interference in LTE-U/Wi-Fi coexistence and show improvements on the overall throughput.
    \item \textbf{Channel selection:} In scenarios where spectrum slices are not possible, a solution can be using the channel selection algorithms to search for clean channels~\cite{qualcomm_tech}. In~\cite{Al-dulaimi2015}, the authors propose an adaptive LBT mechanism that incessantly switches between the channels, impeding the channels for being occupied for a long time. Adopting a priority access approach, authors of~\cite{Song2015} set the LTE-U with a higher priority than Wi-Fi to access the channel and demonstrated that with this approach, the coexistence is enhanced since LTE-U is more robust to the interference. Another way is to limit the LTE presence to increase the chance of Wi-Fi transmitting, as presented in~\cite{Nihtil2013}.
\end{itemize}

Most of the techniques mentioned so far were all non-cooperative since they do not consider the cooperation of coexisting technologies to mitigate interference. However, the coexistence problem presents attributes that can enable the application of cooperative techniques. The authors of~\cite{Al-dulaimi2015} proposed a centralized cooperative control management to employ the virtualization of some network functions, allowing the continuous transfer of resources between unlicensed technologies using cloud control of the distributed access points. In~\cite{paper_erika_2013,paper_fuad_MegMEow}, the proposed solution has networks with modes of operation in which messages are exchanged between them to negotiate which mode the networks should operate and which parameters to use to enable a better coexistence. If a coexisting system is detected, one synchronization phase is started, and consequently, another negotiation phase starts to choose the coexistence parameters. 

%%%% FIM DA SEGUNDA PARTE %%%%

%%% INCIO DA TERCEIRA PARTE %%%%

As previously mentioned, for regions where LBT is mandatory and transmissions are limited to burst, coexistence can be enabled by disabling the standard LTE transmissions on specific subframes using the almost blank subframe (ABS) LTE functionality~\cite{paper_erika_2013}. In~\cite{Cano2015}, the authors solve the problem of coexistence by optimizing the probability of access to the channel. They proposed a proportional fair allocation scheme based on equal distribution of channel access among competing entities, also considering idle periods, successful transmissions and collisions for Wi-Fi networks. In turn, authors of~\cite{p_eval_globecom} leveraged the time division duplex (TDD) functionality of LTE to support various numbers of uplink and downlink frames to grant the fair coexistence between LTE-LAA and Wi-Fi. The authors of~\cite{Xing2015} proposed another technique to adapt the number of subframes according to the traffic load from WLAN systems. Yet considering LAA/Wi-Fi coexistence, the work \cite{Choi2015} presents an investigation of offered transmission time of LTE-LAA to evaluate its performance compared to Wi-Fi.

Regarding the use of machine learning techniques to improve the coexistence of access technologies in the unlicensed band, the authors of~\cite{Ayoubi2018,Zhou2019} survey use cases and appropriate techniques that can be leveraged for wireless communication problems in general. In~\cite{Galanopoulos2016}, a reinforcement learning technique called double Q-Learning for LAA and Wi-Fi coexistence is proposed. This solution is based on discontinuous transmissions (DTX) and transmit power control (TPC) to accurately select the channel, which leads to a better coexistence of LAA and Wi-Fi, as stated by the authors. Also, the authors of~\cite{Castane2017} proposed two solutions, one with Q-Learning and other with game theory, to tackle the problem of channel selection in an indoor scenario. Both solutions indeed enable a better coexistence, providing improvements in data rate. The work~\cite{Rupasinghe2015} proposes an algorithm to dynamically select the LTE-U duty cycle ratio in coexistence with Wi-Fi. The results demonstrate that system capacity increased for simulations with file transfer protocol (FTP) data traffic. Still using Q-Learning, an algorithm applied in a multichannel scenario has the mission of dynamically choosing the least populated channel and also the duty cycle value of that channel to improve system performance ~\cite{Yu2019}. Two other algorithms also use the dynamic choice of duty cycle to achieve fairness in coexistence, differing only in the metric used~\cite{Maglogiannis2018, Bajracharya2019}. 

%%%% FIM DA TERCEIRA PARTE 

In our previous contribution~\cite{Dm_csat}, we proposed a reinforcement learning framework based on Q-Learning applied to LTE-U to automatically adapt the duty cycle parameter to deal with the interference profile between the coexisting systems. However, the evaluated scenario presents only one device and one cell for each operator. So, to better evaluate the coexistence scenario between these two technologies with the proposed framework, as well as to extend our previous conclusions, we proposed herein a multi-cell evaluation, with even more devices (and cells) so that the interference profile can be more realistic and more challenging to track. Thus, in this paper, we highlight the following contributions:

\begin{itemize}[leftmargin=*,labelsep=5.8mm]
    \item  A study of the coexistence problem between LTE-U and Wi-Fi in a multi-cell scenario with co-channel and inter-radio access network (RAT) interference under changing offered data rate that better approximates a possible cellular broadband IoT deployment. This evaluation differentiates from the previous one~\cite{Dm_csat} due to the inclusion of multiples users and multiple interfering RATs.
    \item A new centralized Q-Learning mechanism to dynamically adjust the duty cycle pattern so that it can achieve higher aggregated data rates per user and operator. This centralized mechanism differentiates from the previous one proposed in~\cite{Dm_csat} as it operates under co-channel interference deployment, while the previous one, which is embedded only in a single operator, only considers a single source of interference. Furthermore, the modeling of the proposed solution in~\cite{Dm_csat} takes into account the operator's data rates being the same as the user's data rate, since there is only one user per operator, whilst the modeling of the proposed solution in this current work discriminates between these data rates. Also, there is no coordination in the previous mechanism, consequently there is no communication among cells (base stations).
    \item Extend evaluations and conclusions regarding LTE-U/Wi-Fi coexistence, without and with the proposed framework, to the multi-cell scenario with changing offered data rate and more challenging interference profile.
\end{itemize}

\section{Wi-Fi and LTE-U MAC Layer}
\label{review}
Medium access control is the enabler for the communication of multiple stations in a network since it provides channel access control, besides addressing and power control. In a coexistence scenario, each operator can use a different mechanism to control the medium access. So, it is imperative to review the basics of Wi-Fi and LTE-U and show their differences. In this section, we review the 802.11 MAC layer, followed by some specifications of LTE-U. 

\subsection{The 802.11 MAC Layer}

Since the first release of 802.11 standards in 1997, several enhancements have been developed in the physical layer targeting higher data rates, besides amendments to support additional features. But even with these enhancements, the MAC layer of the following standards still share the same primary MAC mechanism, which is primarily based on the distributed coordination function (DCF). This function defines how the 802.11 devices will access the wireless medium, described as follows.

The DCF is classified as an LBT contention-based mechanism and operates in a distributed manner without any coordination. Whenever a station (STA) wishes to transmit, it has to first perform a clear channel assessment (CCA) for a fixed duration of time called distributed inter-frame spacing (DIFS)~\cite{schiller2003mobile}. If the medium stays idle for DIFS, then the station can take control of the medium and starts to transmit for a period called channel occupation time (CoT). As the station transmits, it controls the channel by keeping a gap between each sent frame of one sequence, the short inter-frame spacing (SIFS), that is the necessary time for the transmitter to receive an acknowledgment (ACK) frame indicating that the receiver successfully received the frame. To avoid other stations accessing the channel duration of the transmission, SIFS is always smaller than DIFS, so that stations do not mislead a SIFS gap with the channel being completely idle. But if the channel goes busy before DIFS is completed, the station defers for other DIFS plus a random backoff time. If the medium remains idle for a DIFS duration plus the backoff time, the STA determines that it can take control of the channel and begins to transmit~\cite{book_wifi}.  

A random backoff time in the DCF approach provides the collision avoidance feature, since whenever the channel transitions from busy to idle, it is possible that multiple STAs were waiting and may be ready to transmit simultaneously. Thus, allowing each station to select a random time decreases the probability of collisions when the channel goes idle. To initiate a random backoff procedure, a station waits for the channel to be idle for a DIFS, then it picks an integer random backoff count in the [0, CW] range, where CW---contention window. 

%Figure~\ref{fig:wifi_back} illustrates the backoff procedure in a Wi-Fi transmission. The backoff counter decreases for that select number for each idle slot time and once it reaches zero, and the channel remains idle, the station transmits immediately~\cite{book_wifi}. If the channel goes busy while the backoff procedure is on the progress, the counter is suspended and only returns to decrease if the channel goes idle for DIFS again. 

%\begin{figure}[H]
 % \centering
 % \includegraphics[width=0.9\linewidth]{figs/wifi_scheme.png}
 % \caption{Wi-Fi transmission procedure with backoff.}
 % \label{fig:wifi_back}
%\end{figure}

The CW value, used in the backoff procedure, always starts with a lower bound $CW_{min}$ and every time that happens an unsuccessful transmission, the value of CW is updated according to $CW=[2\cdot(CW+1)-1]$,  %please confirm it is letter x or mulplication sign. 
% It a multiplication. We change by \cdot.
with $CW_{max}$ being the upper bound. Once  a successful transmission occurs, CW is reset to the lower bound $CW_{min}$~\cite{book_wifi}.

DCF uses two types of carrier sense functions to check the state of the medium: 

\begin{itemize}[leftmargin=*,labelsep=5.8mm]

\item    \textbf{Physical (PHY) carrier sense}: Two PHY %Please define PHY if appropriate. ok, included.
functions based on preamble detection (from Wi-Fi transmissions) and energy (from all transmissions in the bandwidth) to check the channel's state. By measuring the energy level of the channel and comparing if the level is above a certain threshold, these functions can assure the medium is not busy~\cite{ieee_specs}.
\item    \textbf{Virtual carrier sense}: A MAC function based on one field of the MAC header with the expected transmission duration. All stations that receive the information set their network allocation vector (NAV) and wait still this time had passed to start sensing the channel again~\cite{ieee_specs}.  
\end{itemize}

Important changes are provided in other 802.11 related products, such as: a 60~GHz carrier PHY and a scheduled-fashion MAC access of the IEEE 802.11~ad~\cite{norma80211ad,normaWiGIG} (for broadband access with limited coverage); a modified MAC to provide a higher number of connected devices of the 802.11~ad~\cite{park_ah_2015} (achieving a trade-off between the goals of the wireless personal area network (WPAN) and the low-power wide-area network (LPWAN)); and the improvements of dual-band (2.4 and 5~GHz) operation of the IEEE 802.11~ax (Wi-Fi~6)~\cite{Khorov2019,PAR_802_ax}, enabling bandwidth-hungry applications in a dense deployment scenario (4K/8K video, virtual and augmented reality (VR/AR) applications).

\subsection{The LTE-U Coexistence Mechanisms}

\begin{figure}[H]
  \centering
  \includegraphics[width=0.85\linewidth]{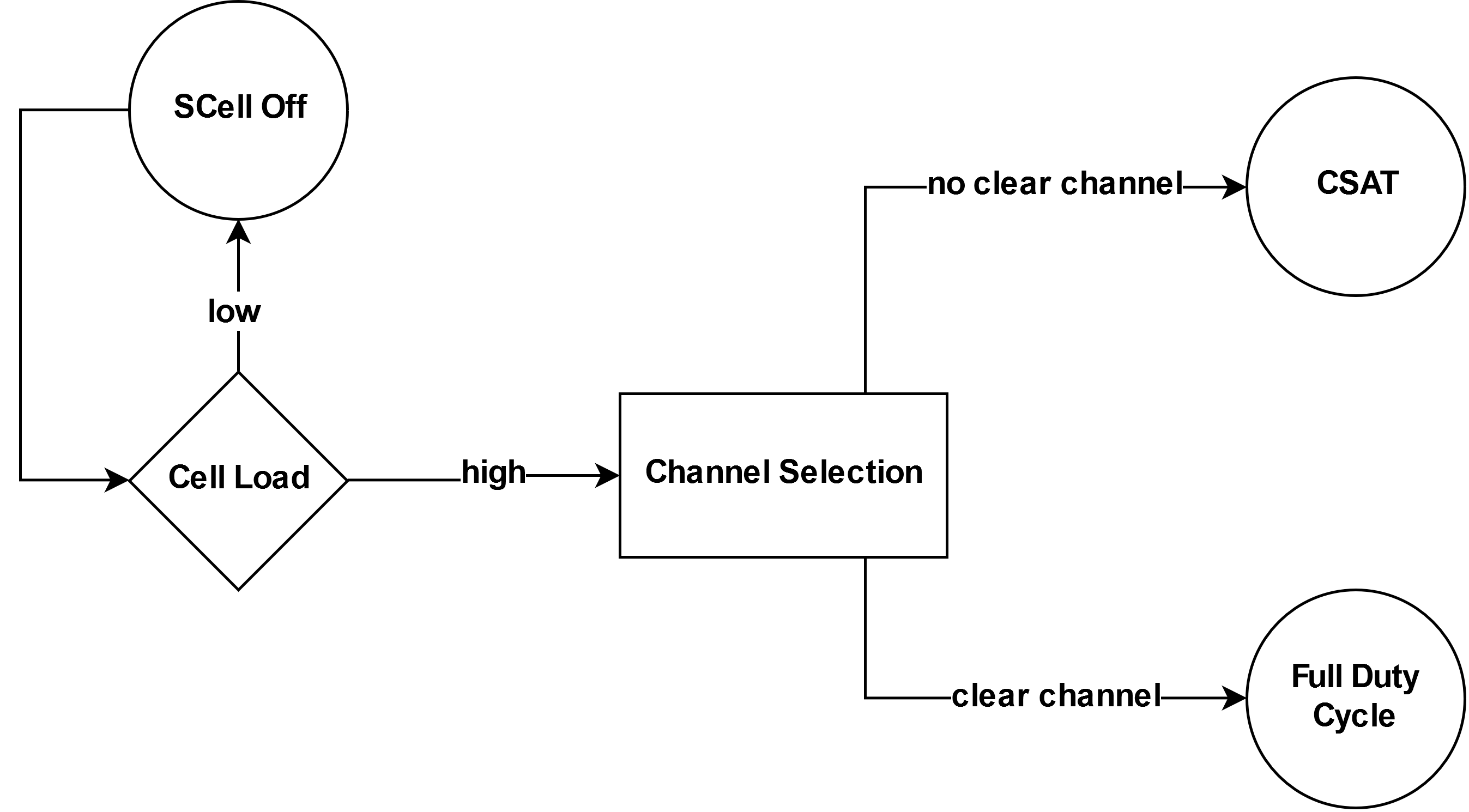}
  \caption{Coexistence flowchart of long-term evolution-unlicensed (LTE-U) adapted from \cite{qualcomm_tech}.}
  \label{fig:lte_u_flow}
\end{figure}

LTE-U is underpinned by three mechanisms to enable coexistence with Wi-Fi and other LBT-like technologies in the unlicensed band: supplemental downlink (SDL), channel selection and carrier-sense adaptive transmission (CSAT)~\cite{qualcomm_tech}. These mechanisms are carefully designed by software and allow the LTE deployment compatible with 3GPP Releases~10~and~11~\cite{qualcomm_tech}. Figure~\ref{fig:lte_u_flow} presents the relational flowchart of the three mechanisms that enable LTE-U for coexistence. 

\hl{\textbf{Supplemental downlink (SDL)}} %please confirm whether the bold is necessary. Yes, it is.
was proposed in 3GGP Release 9 to enable the bonding of paired and unpaired spectrum bands for enhancing downlink transmissions over single-cell deployments~\cite{qualcomm_sdl}. This feature enables the use of one unpaired band to create a downlink-only carrier besides the traditional downlink/uplink frequency division duplex (FDD) scheme in the paired spectrum. Release 10 introduced the possibility of bonding up to three carriers utilizing carrier aggregation that is another feature introduced in this release. SDL with carrier aggregation allows the transmission of data to a station using more than one downlink carrier and presents a division into the primary cell (PCell) carrier and secondary cell (SCell) carrier. For LTE-U, PCell is anchored in a licensed band and carries control and signaling information, while SCell is anchored in the unlicensed band and is used to boost downlink capacity opportunistically. Depending on the traffic load, the LTE base station (eNodeB) can turn the SCell on or off. If the load is light, the eNodeB can use only the PCell and turns off the secondary one, thus mitigating the interference in the unlicensed band. But if the load is heavy, the SCell is enabled and operates following the channel selection coexistence mechanism. 

\textbf{Channel selection} enables the LTE cell to choose the cleanest channel for the secondary cell based on measurements of LTE and Wi-Fi coexisting technologies. It provides a behavior where LTE-U always tries to avoid interference, choosing the channel with the lowest interference profile, releasing the remaining channels to Wi-Fi or other LTE-U base stations. Channel selection operates in an on-going base, then every time the transmitting station finds a cleaner channel, it begins an operation of switching to the cleaner channel. The interference measurements are based on energy detection since it does not discriminate the types of interfering sources. The interference measurement  can also be technology-specific based on the detection of Wi-Fi preambles or LTE-U synchronization signals. Furthermore, for some LTE-U deployments, the channel selection mechanism already satisfies coexistence requirements~\cite{qualcomm_tech}.       

\begin{figure}[H]
  \centering
  \includegraphics[width=0.75\linewidth]{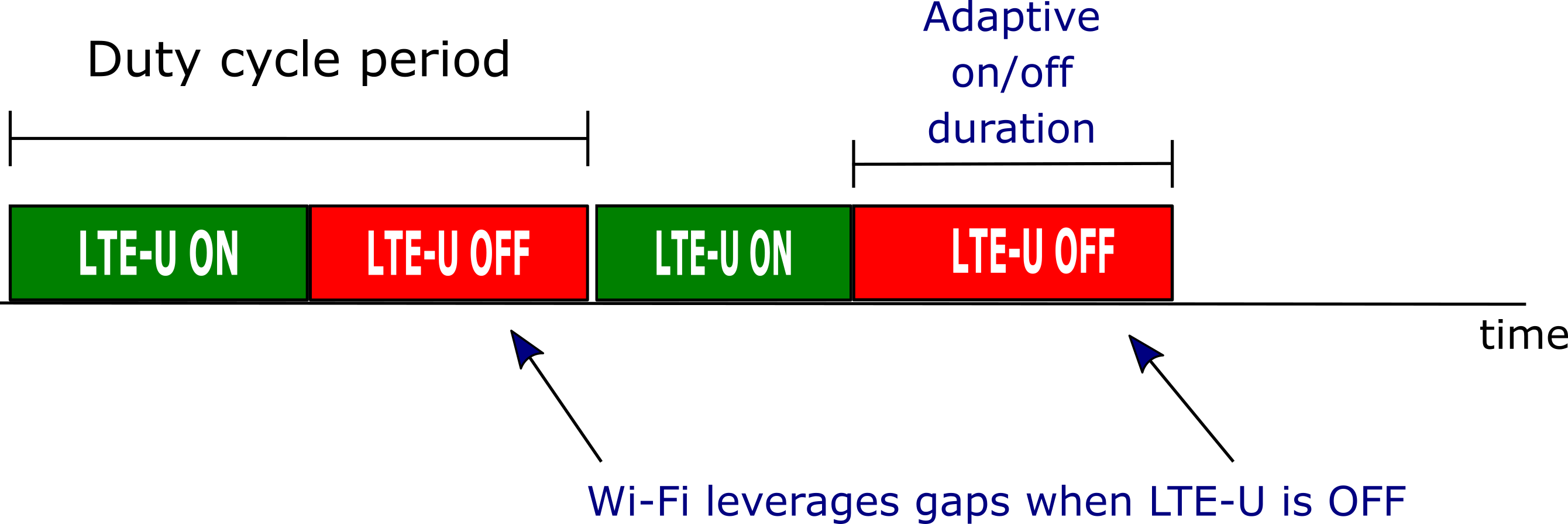}
  \caption{LTE duty cycle approach.}
  \label{fig:lte_dc}
\end{figure}

If channel selection does not find a clear channel, the \textbf{carrier-sense adaptive transmission (CSAT)} algorithm is used to enable the coexistence. CSAT is a time division multiplexing (TDM) algorithm based on long-term medium sensing. Before any attempt of transmission, the CSAT algorithm senses the medium for a long duration than the other coexisting technologies, and according to the observed medium activity, it proportionally adopts a duty cycle behavior activating and deactivating the SCell, as illustrated in Figure~\ref{fig:lte_dc}. 

During the activated periods, LTE-U transmits as a standard LTE, and during the deactivated period, the channel is released so that the coexisting technologies can transmit. LTE-U operating with CSAT algorithm is also called, in literature, as LTE-DC~\cite{Dm_csat,laa_pedro, Santana2017}. As Qualcomm mainly proposed CSAT, the details of its algorithm implementation are hidden, opening up the possibility of alternative implementations by the industrial and academic community.

After a successful CSAT procedure, the duty cycle behavior in LTE-U can be performed by the almost blank subframe (ABS) functionality. This solution presented in~\cite{paper_erika_2013}, following the already defined feature with the same name presented in Release 10, but in the context of heterogeneous networks (HetNets). In LTE HetNets, cells present different sizes (macro, micro, pico and femto). Taking into account the essential ABS operation in this context, during the transmission time of the macro cell, the data transmission only occurs in individual subframes. Thus the remaining subframes can be used by the other cells. In the context of LTE-U, the data can be transmitted in the defined subframes while Wi-Fi, for example,  could exploit the blanked LTE-U subframes to perform its transmission. The proportion of subframes to blank out is defined by the duty cycle ratio informed by the CSAT algorithm after its medium sensing procedure.

\section{System model, Evaluation Scenario and Preliminary Results}
\label{model}
Before any attempt to incorporate the proposed reinforcement learning framework into the coexistence problem, the evaluation scenario must be defined, and the preliminary results must show why it is possible and necessary to use the proposed framework. This section presents the scenario as well as the preliminary results without the proposed reinforcement learning algorithm.   

\subsection{System Model and Evaluation Scenario}

\begin{figure}[H]
  \centering
  \includegraphics[width=0.7\linewidth]{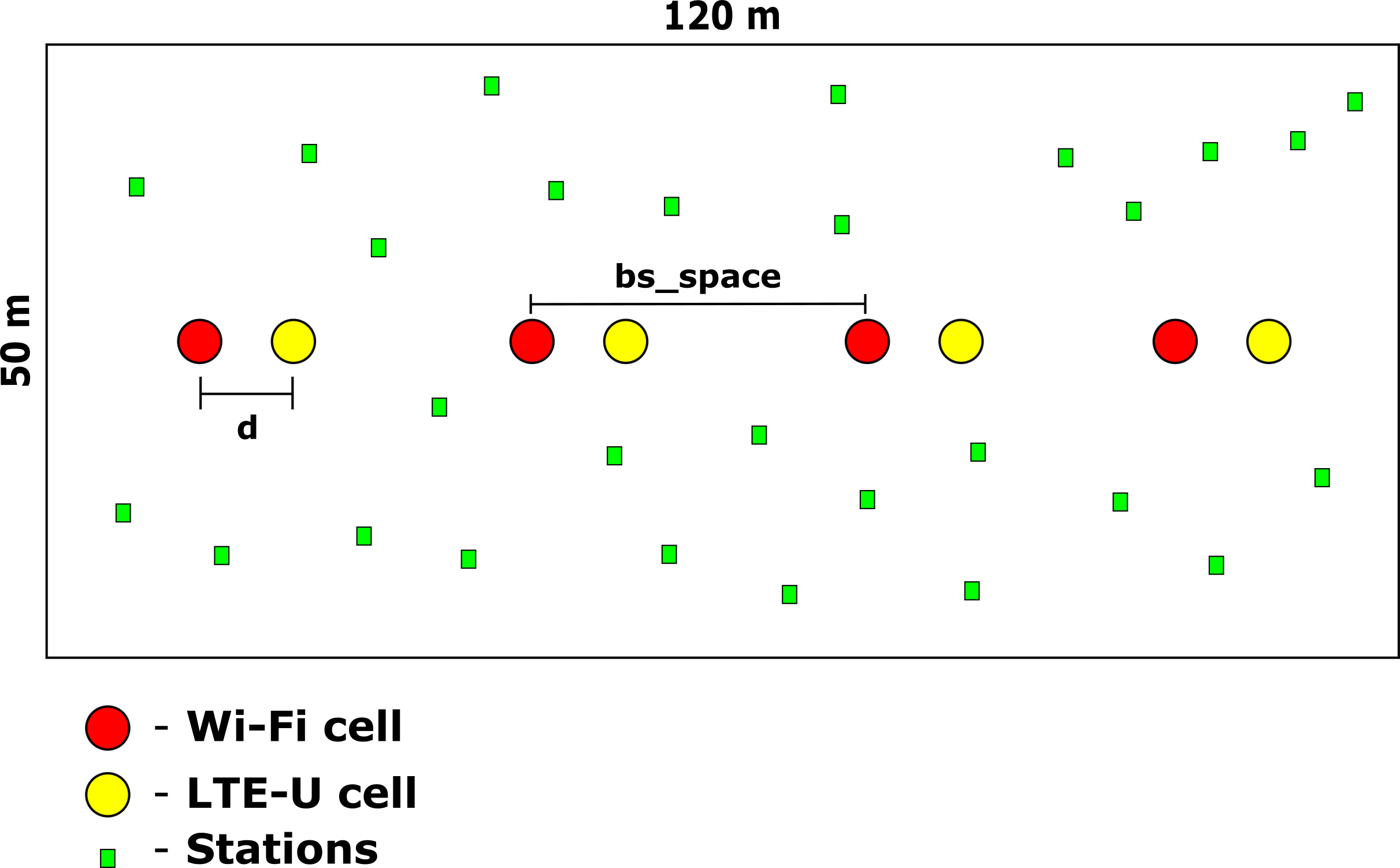}
  \caption{Evaluation indoor scenario proposed by the 3rd generation partnership project (3GPP).}
  \label{fig:scenario}
\end{figure}

To better evaluate the coexistence problem between Wi-Fi and LTE-U, 3GPP proposed evaluation scenarios in the technical report TR36889~\cite{TR36_889}. Among the proposed scenarios, the indoor (Hotspot) scenario is the most challenging and complex since it comprises not only multiple users but also interference between the same and different access technologies. Figure~\ref{fig:scenario} presents the 3GPP indoor scenario considered in this work. This scenario is composed of two operators, called operator A and operator B, in an unlicensed band deploying four small cells each. The deployment is on a building room with dimensions of 50 $\times$ 120 m, no walls, and with the four base stations of each operator equally spaced in the X-axis defined by the $d$ \hl{and} $bs\_space$  %please confirm whether bold is necessary. No, it better in math mode.
distances. Besides the four cells, forty stations are randomly distributed in this rectangular region with no mobility~\cite{TR36_889}. 

We use the ns-3 simulator~\cite{ns3_repository} to model the Wi-Fi, the LTE-U and their coexistence. The ns-3 is an open-source C++ network simulator based on discrete event simulation for research and educational purposes, also complying with technical norms from standard organizations like IEEE, 3GPP and Wi-Fi Alliance. As part of a project funded by Wi-Fi Alliance, the \textit{Centre Tecnològic de Telecomunicaciones de Catalunya} (CTTC) and the University of Washington cooperated and created a branch of this simulator with the system modeling and coexistence evaluation of LTE-U and Wi-Fi, named as ns3-lbt~\cite{ns3_lbt_tech_description}. In this version, the implemented duty cycled transmission of LTE-U follows the approach presented in~\cite{paper_erika_2013}, which is based on LTE ABS functionality, where the ABS pattern is defined with the duration of 40~ms. Since each subframe of an LTE transmission lasts for 1 ms, one bitmask of 40 values is defined (for each subframe) to control the ON--OFF pattern. Both the core and discussed modifications of this simulator are found on \textit{http://code.nsnam.org/laa/ns-3-lbt/file/9529febb7ebc}.

Since the evaluation scenario presents a challenging interference profile, simulations were performed for different offered data rates for each duty cycle to map possible patterns. Following this approach, the DC value used is in the range of 0.2 to 0.9 with a granularity 0.1. The offered data rate follows the full buffer scheme, i.e., user datagram protocol (UDP) transmission with a constant bit rate, since we want to analyze the situation where there is always data to transmit, thus allowing an intense dispute to access the medium. This UDP offered data rate represents the available data rate capacity for each station, and a variable called \textit{UDPRate} controls it. The parameters used on the simulation, for both LTE-U and Wi-Fi, are presented in Table~\ref{table_parameters_simu}. The simulation is set to run for 20 seconds since it is a sufficient duration that is possible to see the relation between the offered data rate and duty-cycle values in this scenario.

\begin{table}[H]
\caption{Simulation parameters.}
\label{table_parameters_simu}
\centering
\tablesize{\normalsize}
\begin{tabular}{ l l }
\toprule
\multicolumn{1}{c}{\textbf{Wi-Fi parameter (802.11n-HT PHY/MAC)}} & \multicolumn{1}{c}{\textbf{}}  \\
Bandwidth                                                         & 20 MHz  \\
CCA---Energy Detection threshold                                  &$ - $62 dBm     \\
CCA---Carrier sense threshold                                     & $- $82 dBm \\
Bit Error Rate (BER) target                                       & $10^{-6}$ \\
\midrule
\textbf{LTE parameters}                                           & \\
Bandwidth                                                         & 20 MHz  \\
Packet scheduler                                                  & Proportional fair  \\
ABS pattern duration                                              & 40 ms  \\
Duty cycle values                                              & \{0.2, 0.3, 0.4, 0.5, 0.6, 0.7, 0.8, 0.9\} \\

\midrule
\textbf{Common parameters}                                        & \\
Tx power                                                          & $- $18 dBm \\
Traffic model                                                     & UDP full buffer \\
Mobility                                                          & Constant position  \\
\midrule
\textbf{Scenario}                                        &  \\
d                                                          & 5~m\\
bs\_space                                                     & 25~m \\
Number of LTE-U APs                                                     & 4 \\
Number of LTE-U stations                                                & 20 \\
Number of Wi-Fi APs                                                     & 4 \\
Number of Wi-Fi stations                                                & 20\\
Path loss and Shadowing                                                & ITU InH            \\
Cell selection criteria                                                & For Wi-Fi, AP with strongest RSS.  \\
  & For LTE-U, cell with the strongest RSRP.  \\
UDPRate  & \{2 Mbps, 4 Mbps\}.  \\
\bottomrule
\end{tabular}
\end{table}

\subsection{Preliminaries Results}

\begin{figure}[H]
  \centering
  \includegraphics[width=0.6\linewidth]{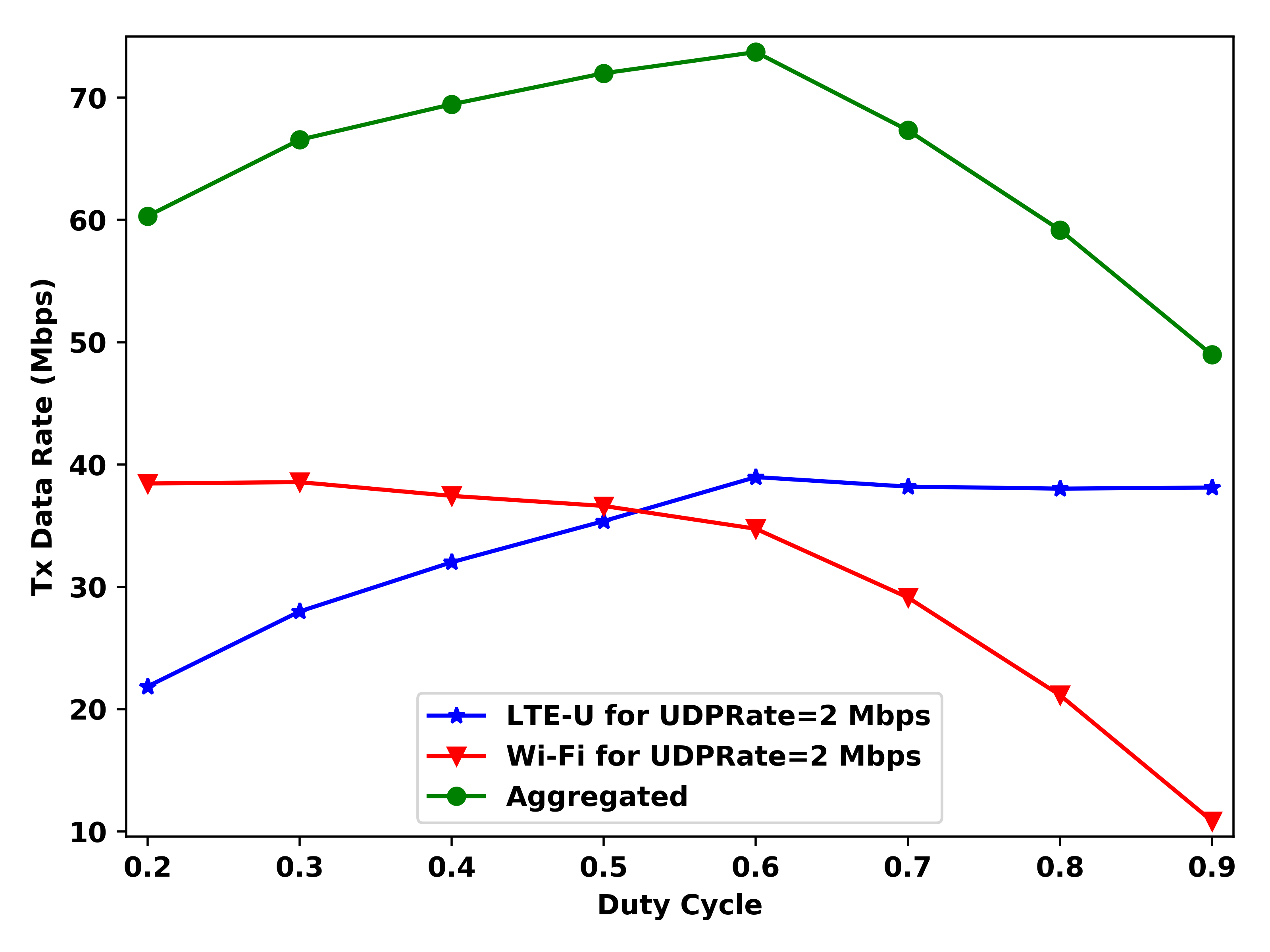}
  \caption{Data rate vs duty cycle (DC) for 2~Mbps of offered data rate per station.}
  \label{fig:static2}
\end{figure}

Figure~\ref{fig:static2} presents 
%a line graph showing 
the relation between the transmitted data rate (per operator) versus duty cycle value for $UDPRate=2$~Mbps. In theory, the expected transmitted data rate for each operator is the sum of its user's rates (without interference), then it would be expected for each operator $20 \cdot 2=40$~Mbps. 

 And as we can see, this theoretical data is nearly reached for each operator for some DC values, but never at the same time. For this behavior, the aggregated transmitted data rate (LTE-U + Wi-Fi) presents its maximum only for DC 0.6, even though this DC value does not present the best possible data rate for Wi-Fi. This duty cycle value indicates that for this configuration and the offered data rate, the system reaches its maximum rate with LTE-U having more channel access time than Wi-Fi. Furthermore, the LTE-U's data rates for the DC values from 0.6 and on do not present meaningful variations, indicating that for this \textit{UDPRate} as the DC value keeps increasing, the most affect access technology is the Wi-Fi. The same pattern also occurs for Wi-Fi for DC values less than 0.5. 

% \begin{figure}[H]
%   \centering
%   \includegraphics[width=0.8\linewidth]{figs/1Mbps_static.png}
%   \caption{Data rate vs DC for 1Mbps of offered data rate per station.}
%   \label{fig:static1}
% \end{figure}

\begin{figure}[H]
  \centering
  \includegraphics[width=0.6\linewidth]{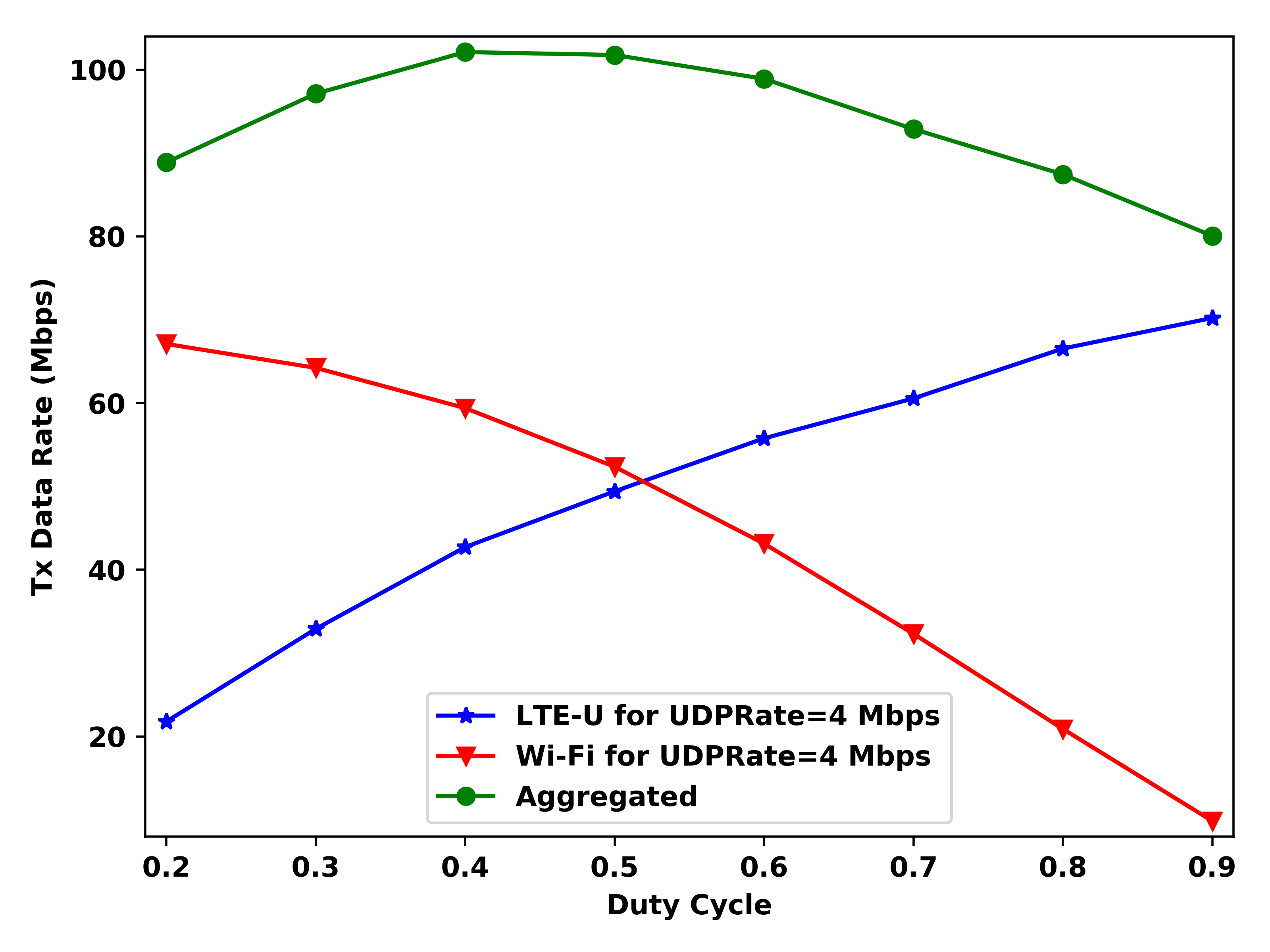}
  \caption{Data rate vs DC for 4~Mbps of offered data rate per station.}
  \label{fig:static4}
\end{figure}

Figure~\ref{fig:static4} is the similar to Figure~\ref{fig:static2}, but now with $UDPRate = 4$~Mbps. For this offered data rate, the maximum aggregated transmitted data rate is reached for a DC value of 0.4., indicating that the system, for this situation, reaches its maximum with Wi-Fi having more channel access time than LTE-U. Furthermore, for this \textit{UDPRate}, the transmitted data rate for each operator is more affected by the DC value. While for $UDPRate=2$~Mbps, and duty cycle values bigger than 0.6, the LTE-U's transmitted rate slightly changes, now for $UDPRate=4$~Mbps, the data rates keep increasing with the DC value, and the Wi-Fi ones keep decreasing. 

%As it can be noted, 
As a general result for the multi-cell environment, the duty cycle value has a significant influence on the transmitted data rates combined with the \textit{UDPRate} used. This pattern also occurs in other scenarios, and different data rates, as presented in~\cite{Dm_csat} for a single-cell scenario, showing a general pattern in this LTE-U/Wi-Fi coexistence problem. This problem also gets more complicated if the \textit{UDPRate} changes over time, as the best DC value also changes per each offered data rate. Then, for setting the goal for maximizing the system throughput in a dynamic offered data rate scenario, it is necessary to use a tool for automatically choosing the optimal DC value on the fly. The next section presents our proposed reinforcement learning framework based on Q-Learning to address this~problem.

\section{Proposed Reinforcement Learning Framework}
\label{frame}

%In the problem of LTE-U/Wi-Fi coexistence presented, we saw that each DC value maps into a different operator's and aggregated transmitted data rate. 
If we set the goal for maximizing the aggregated transmitted data rate as the time passes, and the offered data rate changes, a new DC value must be chosen to keep the performance on the maximum point, then bringing the problem to a sort of sequential decision-making problem. In this decision-making problem, each new chosen duty cycle value leads the system to a new state (a new value of aggregated transmitted data rate) that can change if the offered data rate also changes. So, for choosing the best duty cycle value for each case, the correlated decision-making problem must be solved. We address the Q-Learning as a solution to this problem.

\subsection{Q-Learning}
\begin{figure}[H]
  \centering
  \includegraphics[width=0.35\linewidth]{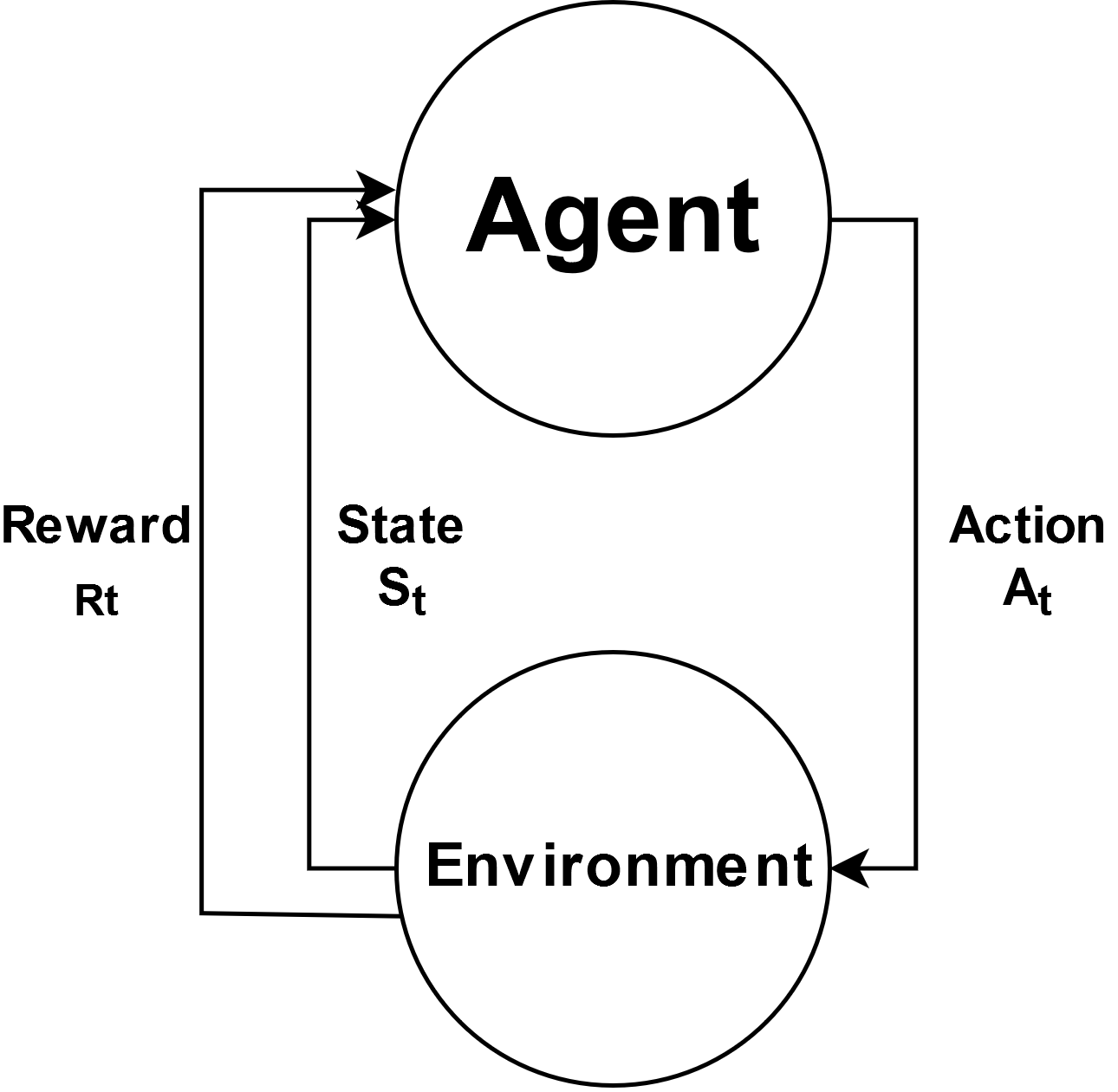}
  \caption{Agent--Environment interaction in a Markov decision process (MDP) adapted from~\cite{sutton}.}
  \label{fig:reinfor}
\end{figure}

Q-Learning is a reinforcement learning algorithm proposed in~\cite{Watkins1992} that combines dynamic programming and Monte Carlo concepts to solve sequential decision-making problems in the form of Markov decision processes (MDP)~\cite{sutton}. In MDP, we have a decision-maker that is called an agent, and all other things are called the environment. As the agent interacts with the environment, it chooses actions, and in turn, the environment responds to these actions in the form of observations or new situations. This observation returned from the environment are called rewards and are numerical values that the agent uses to achieve a goal, i.e., maximizing the sum of all received rewards. Figure \ref{fig:reinfor} presents the relationship of these concepts in a diagram.  

Practically all reinforcement learning algorithms involve the estimation of a \textit{value function}~\cite{sutton}. The value function is a map from state--action pairs to a value that estimates how good or bad is a given state for the agent to be in. This notion of ``how good'' is defined in terms of the immediate and all expected future rewards that an agent receives as the learning process occurs, and since the future rewards depend on actions taken, a track of actions defines a policy $\pi$. Formally, a policy is a mapping from states to actions, and it is changed as a result of the agent's learning process. 

If at time \textbf{t} the action selected \hl{is} $a$,  %a is not bold in the equation. please keep them consistent in the main text. ok, we replaced by math mode.
the system is in state \textbf{s} under policy $\pi$, then it is defined the action-value function for policy $\pi$ as

\begin{equation*}
q_{\pi}(a, s) \doteq \mathbb{E}_{\pi} \left[ \sum_{k=0}^{\infty} \gamma^{k}R_{t+k+1} \bigg\vert S_{t} = s, A_{t = a} \right],
\end{equation*}

\noindent where $\mathbb{E}$ is the expected value, and $\gamma$ is the discount factor controlling the relation between immediate and future rewards. Solving this reinforcement learning task means finding a policy that returns the maximum possible reward over the long run. There is always an optimal policy that is better or equal to the other policies~\cite{watkins} denoted $\pi _{*}$. Then, the optimal value for this optimal policy is defined as

\begin{equation*}
q_{*}(a, s) \doteq \underset{\pi }{max}\ q_{\pi}(s,a).
\end{equation*}

However, solving the reinforcement learning task for $q_{*}$ is not trivial, since it requires the transition probabilities associated with the states and actions taken. To solve this problem, Q-Learning defines a Q-value that approximates $q_{*}$  and solves it independent of the policy being followed. This Q-value is defined by 

\begin{equation}
\label{Q_func}
Q(S_{t}, A_{t}) \xleftarrow{} Q(S_{t}, A_{t}) + \alpha \left[ R_{t+1} + \gamma\underset{a}{max}\ Q(S_{t+1}, a) - Q(S_{t}, A_{t}) \right],
\end{equation}

\noindent where $S_{t}$ is the set of states; $A_{t}$ is the set of actions; $\alpha$ is the learning rate, indicates how much of the learning is taken into account to update the Q-value; $\gamma$ is a parameter that controls how much the future rewards influences to update Q. If $\gamma$ is set to zero, the agents learn to take into account only the immediate rewards, while if $\gamma$ is to set to 1, the agent learns using only the possible future rewards. Both $\alpha$ and $\gamma$ are in the range 0 to 1. The learning task of the agent is represented by updating the relation of Equation~(\ref{Q_func}) for every state it reaches paired with the corresponding action. As presented in~\cite{sutton, Watkins1992}, the only requisite for convergence is all pairs (s, a) continue to be updated. Under this assumption, as each pair is visited several times and the corresponding Q value is updated, the algorithm converges with probability 1 to the optimal $q_{*}$~\cite{sutton, Watkins1992}.     

\subsection{Proposed Framework}

In our work, we propose a centralized, coordinated algorithm to maximize the system aggregated transmitted data rate, and in the basis of Q-Learning we formulate the sets of states $\mathcal{S}$ and actions $\mathcal{A}$ as~follows:

\begin{itemize}[leftmargin=*,labelsep=5.8mm]
\item    The duty cycle values that can be chosen as actions are $\mathcal{A} = \{0.2,\ 0.4,\ 0.6,\ 0.8\}$. These duty cycle values are chosen in a coordinated way. The node running the Q-Learning algorithm uses system information to choose the best DC value that is therefore set to all LTE-U cells;

\item    The states represented by the aggregated transmitted data rate, $\mathcal{S} = \{0,\ 1,\ 2,\ 3\}$, are:

    \subitem \textbf{State 0}: \hspace{0.5cm} $0 <\ T_{tx\ wifi} + T_{tx\ lteu}  \leq \frac{1*M}{4}$ Mbps;
    \subitem \textbf{State 1}: \hspace{0.5cm} $\frac{1*M}{4} <\ T_{tx\ wifi} + T_{tx\ lteu}  \leq \frac{2*M}{4}$ Mbps;
    \subitem \textbf{State 2}: \hspace{0.5cm} $\frac{2*M}{4} <\ T_{tx\ wifi} + T_{tx\ lteu}  \leq \frac{3*M}{4}$ Mbps;
    \subitem \textbf{State 3}: \hspace{0.5cm} $\frac{3*M}{4} <\ T_{tx\ wifi} + T_{tx\ lteu}  \leq M$ Mbps;\\
    
\noindent where $T_{tx\ wifi}$ is the sum of the transmitted data rate of all four Wi-Fi cells, and $T_{tx\ lteu}$ is the sum of all four LTE-U cells. Then, the proposed algorithm works on a centralized way operating only on the operator data rate, LTE-U or Wi-Fi, and not on each cell. In this way, the state formulation can be simplified, since the algorithm operates on two operators and not on four cells. Thus, the number of states can be decreased to the only four states presented. Furthermore, defining the algorithm to work in a centralized operation also softens issues related to energy consumption since the algorithm will be operating in one single node in a coordinated way (or in a cloud server);   

\item The reward for each DC value chosen is set to be $T_{tx\ wifi} + T_{tx\ lteu}$ that will be the aggregated transmitted data rate the system will reach for that DC. The general goal is to minimize the cost function $C~=~T_{target}~-~(T_{tx\ wifi}~+~T_{tx\ lteu})$ that can be also viewed as the maximization of the reward $T_{tx\ wifi} + T_{tx\ lteu}$ over time to fulfill the goal. $T_{target}$ is the desired aggregated transmitted data rate and it is set as a parameter. 

\item The goal is set to maximize the system aggregated transmitted data rate.

\end{itemize}

The choice of these DC values is based on, firstly, the trade-off between computational complexity and the number of actions. This is intrinsic to the Q-Learning algorithm since the higher the number of actions and states, the higher the time is taken by the algorithm to converge. The second point is that these values represent four situations, starting with Wi-Fi having more air time than LTE-U, for $DC = 0.2$, to the opposite situation of LTE-U having more air time than Wi-Fi, $DC=0.8$. Thus, after convergence, the best action (DC value) for the current state will be determined by the optimal Q-value, indicating what percentage of air time for each technology is more suitable in that moment for achieving the optimum goal.   

The intervals between states were chosen as equally spaced intervals from 0 to $M$, where $M$ is the maximum aggregated transmitted data rate that the system can achieve. It depends on the parameters of each system and the coexisting environment.  The value of $M$ differentiates from $T_{target}$, as $M$ is the real maximum that the system can achieve by the parameters used without the proposed solution, while $T_{target}$ is the desired maximum we want to achieve using the proposed algorithm.    

Since the proposed solution operates in a centralized and coordinated manner, the new duty cycle value chosen by the algorithm, in each step, must be sent to all LTE-U cells by the central node. Regarding the simulation, as we use a discrete event simulator, the parameter parsing is by functions and occurs in the updating duty cycle phase, incurring in no latency (and overhead) for sending the new duty cycle value from the central node. However, targeting the real deployments, this communication between nodes can be achieved by jointly utilizing the X2 interface and the X2AP services. Defined by 3GPP in~\cite{3GPP_TS36423}, the X2 is the denomination of the interface that connects one eNodeB to another for exchanging signaling messages, and the X2AP is the related protocol. The X2AP provides several functions as mobility management, cell configuration update, error reporting and others, which can be used by our proposed  Q-Learning mechanism for sending the updated DC value with controlled overhead. However, when it comes to the latency issue, additional studies must be performed considering that the latency using X2 interface is highly influenced by the used network topology, transport technology (e.g., optical fiber and copper-base) and capacity of the backhaul link, which can vary the latency from a hundred of microseconds to 20~ms~\cite{comp_latency, comp_latency2}.

\begin{figure}[H]
  \centering
  \includegraphics[width=1.0\linewidth]{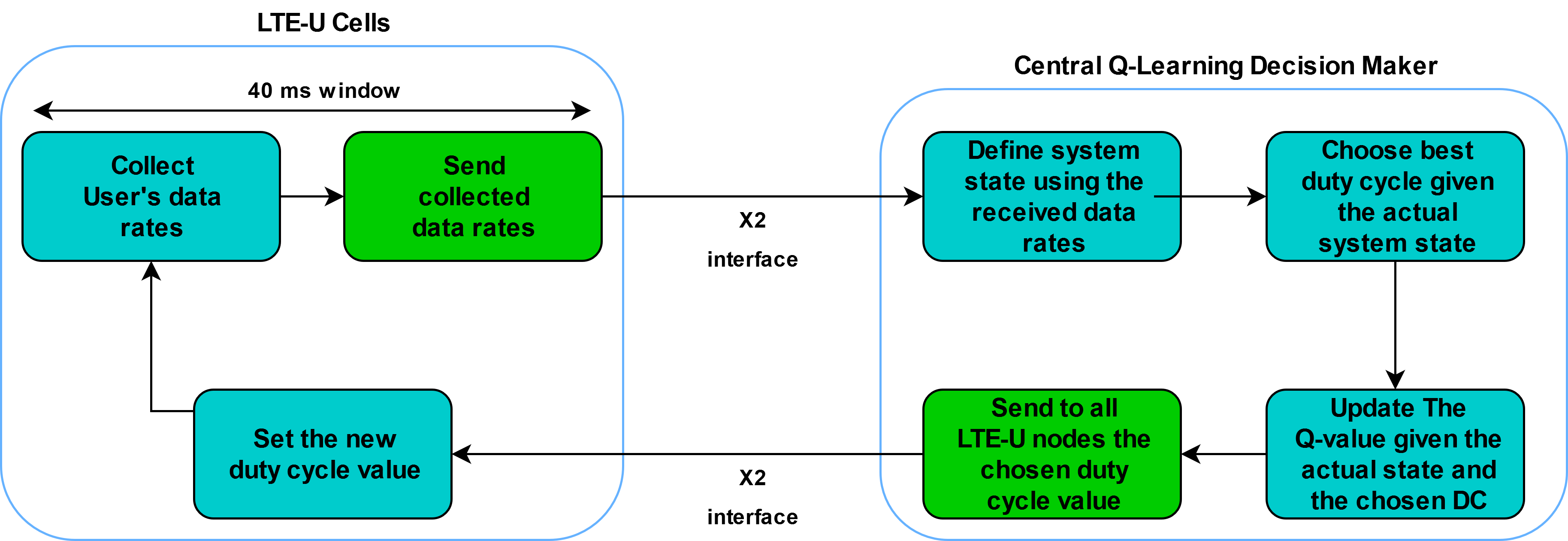}
  \caption{Proposed solution flowchart.}
  \label{fig:workflow}
\end{figure}

Figure~\ref{fig:workflow} shows a flowchart of the proposed solution. From a macro-perspective, the proposed algorithm has two main steps. The first one, at the LTE-U cells, updates the actual DC value and reports the reward (data rates) to the central Q-Learning node. The second step, at the central Q-Learning node, performs a Q-Learning cycle, yielding in a new DC. This new DC value is reported back to the LTE-U cells in which the step one is started again.

 To complement, the pseudo-algorithm of the proposed centralized, the coordinated solution based on Q-Learning is also presented in the Algorithm~\ref{pseudo_qlearning}.\\

\begin{algorithm}[H]
    \SetKwInOut{Input}{Input}
    \SetKwInOut{Output}{Output}

    %\underline{function Euclid} $(a,b)$\;
    %\Input{Two nonnegative integers $a$ and $b$}
		\textbf{Initialize}
		
		\For{$s \in S, a \in A$}
		{
		Start Q-Table with random Q-Values~$Q(s,a)$.
		}
	
		Estimate initial states~$s$.
		
		\textbf{Learning:}
		
		\textbf{Loop}
		
		{
			Generate a random value $r \in U(0,1)$
		}
		
    \eIf{$r < \epsilon$}
      {
        Select an action~$a \in A$ randomly\;
      }
      {
        Select an action $a \in A$ according to $\max\limits_{a}Q(s,a)$\; 
      }
			
		Run $a$\; 
		
		Take the immediate reward~$r$, which is~$T_{tx\ wifi} + T_{tx\ wifi}$\;  
		
		Observe the next state~$s{'} \in S$\; 
		
		Update Q-table according to the expression:
		
		{
		$Q(s,a) \leftarrow (1-\alpha)Q(s,a) + \alpha\left [ r + \gamma \max\limits_{a}Q(s{'},a) \right ]$
		}
		
		$s = s{'}$
		
		\textbf{end loop}
			
    \caption{Q-Learning application for dynamic duty-cycle selection in a multi-cell environment.}\label{pseudo_qlearning}
\end{algorithm}

\section{Proposed Solution Evaluation}
\label{eval}

A new simulation to evaluate the proposed Q-Learning algorithm was performed following the same basic parameters for Wi-Fi and LTE-U presented in Table~\ref{table_parameters_simu}, but with modifications regarding the offered data rate for each station and the total duration of the simulation. The proof-of-conception simulation was performed according to the following scheme:

\begin{itemize}[leftmargin=*,labelsep=5.8mm]
\item    The simulation has a duration of 250 s where the offered data rate changes at specific points as the simulation runs.  
\item    The offered data rate is uniformly picked from $T_{oferred}$ = \{\ 500\ kbps,\ 1\ Mbps,\ 2\ Mbps,\ 4\ Mbps\}.
\item   The specific points in time where the offered data rate changes that occur are uniformly picked from an interval between 10 to 15 s that are then summed to the current time until it reaches the 250 s limit. In this way, during the 250 s simulation, there will be from 16 to 25 changes in data rate (total duration divided by minimum and maximum intervals).
\item The Q-Learning algorithm always operates at the end of the current ABS interval, which is 40~ms, choosing the DC value for the next ABS interval until it converges.
\item To gain statistical confidence, 100 snapshots/repetitions of the 250 s simulation were realized. In each of these snapshots, the stations are uniformly distributed in the scenario. Then, in each snapshot, the stations are in different positions generating different interference profile.
\item Regarding the states, $M$ is set to $160$~Mbps and $T_{target}=M$; \item Regarding the Q-Learning parameters, the values of $\alpha$ and $\gamma$ were chosen as $0.3$ and $0.5$, respectively. These values were the best ones when the algorithm converged in the test phase using the proposed solution.
\end{itemize}

\begin{figure}[H]
  \centering
  \includegraphics[width=0.8\linewidth]{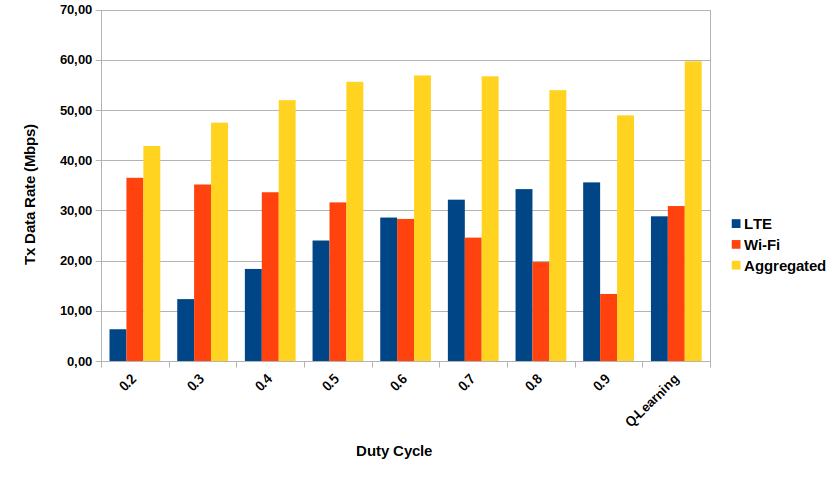}
  \caption{Aggregated transmitted data rate vs DC value including the proposed framework.}
  \label{fig:agg_q_learning}
\end{figure}

Figure~\ref{fig:agg_q_learning} presents a bar graph of the simulation results using some fixed DC values and with our proposed framework. With the proposed framework, the maximum aggregated transmitted data rate is increased when compared with all the duty cycle values without the solution. When compared with the best static results, $DC=0.6$, and $DC=0.7$, the Q-Learning increases the maximum aggregated with around $3$~Mbps more than the static ones, corresponding to a 5\% increase percentage. If we compare the lower \textit{UDPRate} versus the Q-Learning gain, it can correspond to six times increasing (\textit{UDRate} with $500$~kbps). If we focus on each operator, the LTE-U and Wi-Fi transmitted data rates present gains over the ones for $DC=0.6$. However, if we take the other best static DC value, 0.7, only the Wi-Fi presents gain over the static case, even though the system still reaches its maximum. Furthermore, neither the LTE-U data rate with Q-Learning does not reach the maximum it can achieve $DC=0.9$, nor the Wi-Fi at $DC=0.2$. Still, the goal was set to maximize the aggregated $T_{tx\ lteu} + T_{tx\ wifi}$. The algorithm chose the best combination of duty cycle values for each state the system had passed to reach the goal, which led to a combination of data rates for each operator that was not the maximum possible, but the necessary to the goal as it would be expected.      

\begin{figure}[H]
  \centering
  \includegraphics[width=0.7\linewidth]{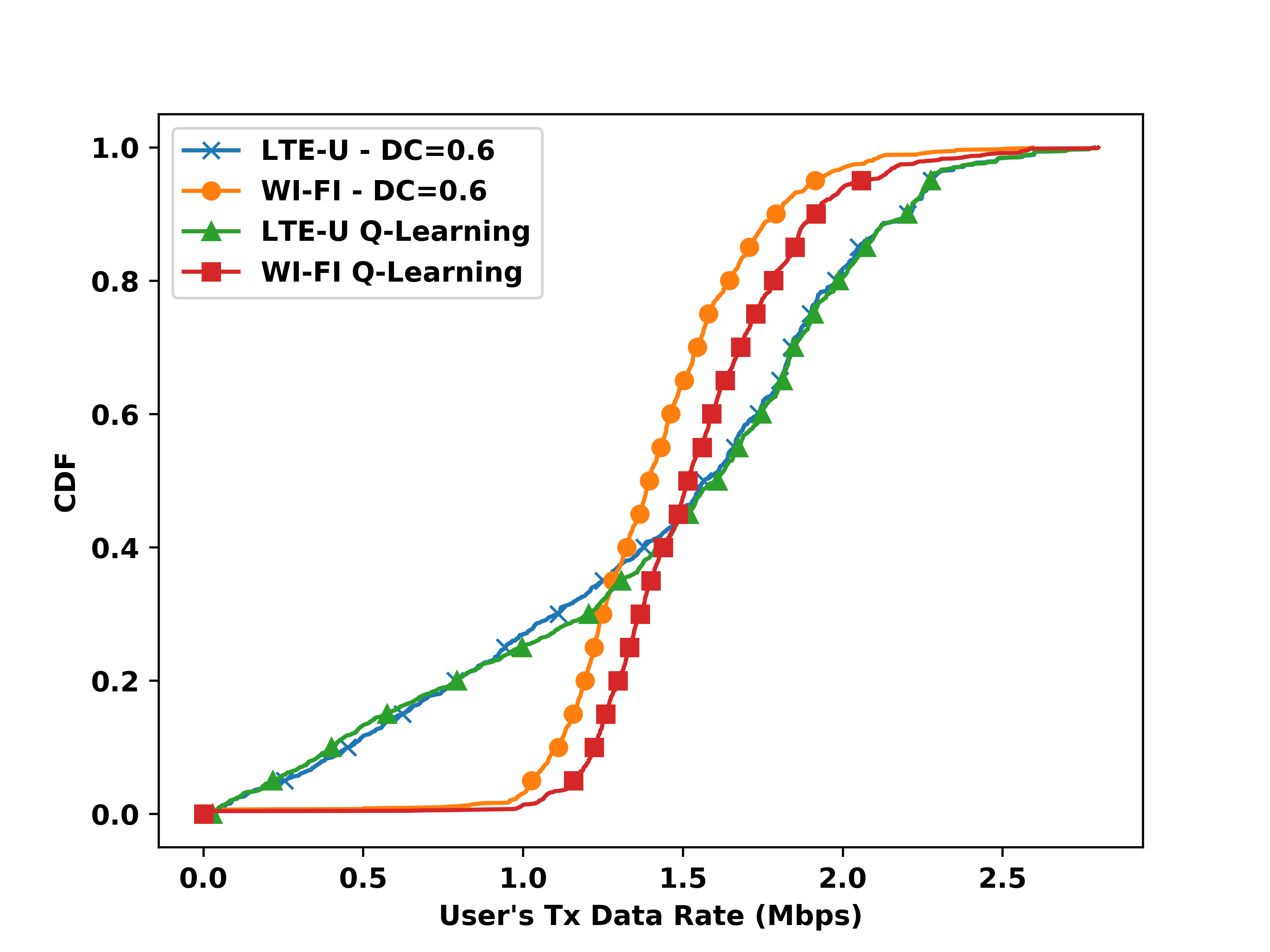}
  \caption{Cumulative density function (CDF) of transmitted data rate per user vs Q-Learning for $DC=0.6$.}
  \label{fig:cdf_q_learning_0.6}
\end{figure}

The results so far have only shown the aggregated transmitted data rate perspective, but the Q-Learning also indirectly operates on each user transmitted data rate. Figure~\ref{fig:cdf_q_learning_0.6} presents a comparison between the cumulative density function (CDF) of the user transmitted data rate for the Q-Learning and the best static duty cycle value 0.6. %As we can see, 
Observing the Q-Learning curves, the transmitted data rate of practically all the Wi-Fi users improved, while all LTE-U users maintained their performance. The successive duty cycle values chose by the Q-Learning algorithm led the system to a behavior where Wi-Fi users have more air time. Consequently, more data rate, than the static $DC=0.6$, but maintaining the mean data rate of LTE-U users, improving the aggregated one. Furthermore, the gain of around $3$ Mbps in the aggregated performance presented in Figure~\ref{fig:agg_q_learning} is higher than the maximum user transmitted data rate for both Wi-Fi and LTE-U that is $2.7$~Mbps. Still, since the maximum data rate is the same with or without the proposed solution, we conclude that the gain in the aggregated data rate is distributed all over the users with less data rate.  

\begin{figure}[H]
  \centering
  \includegraphics[width=0.7\linewidth]{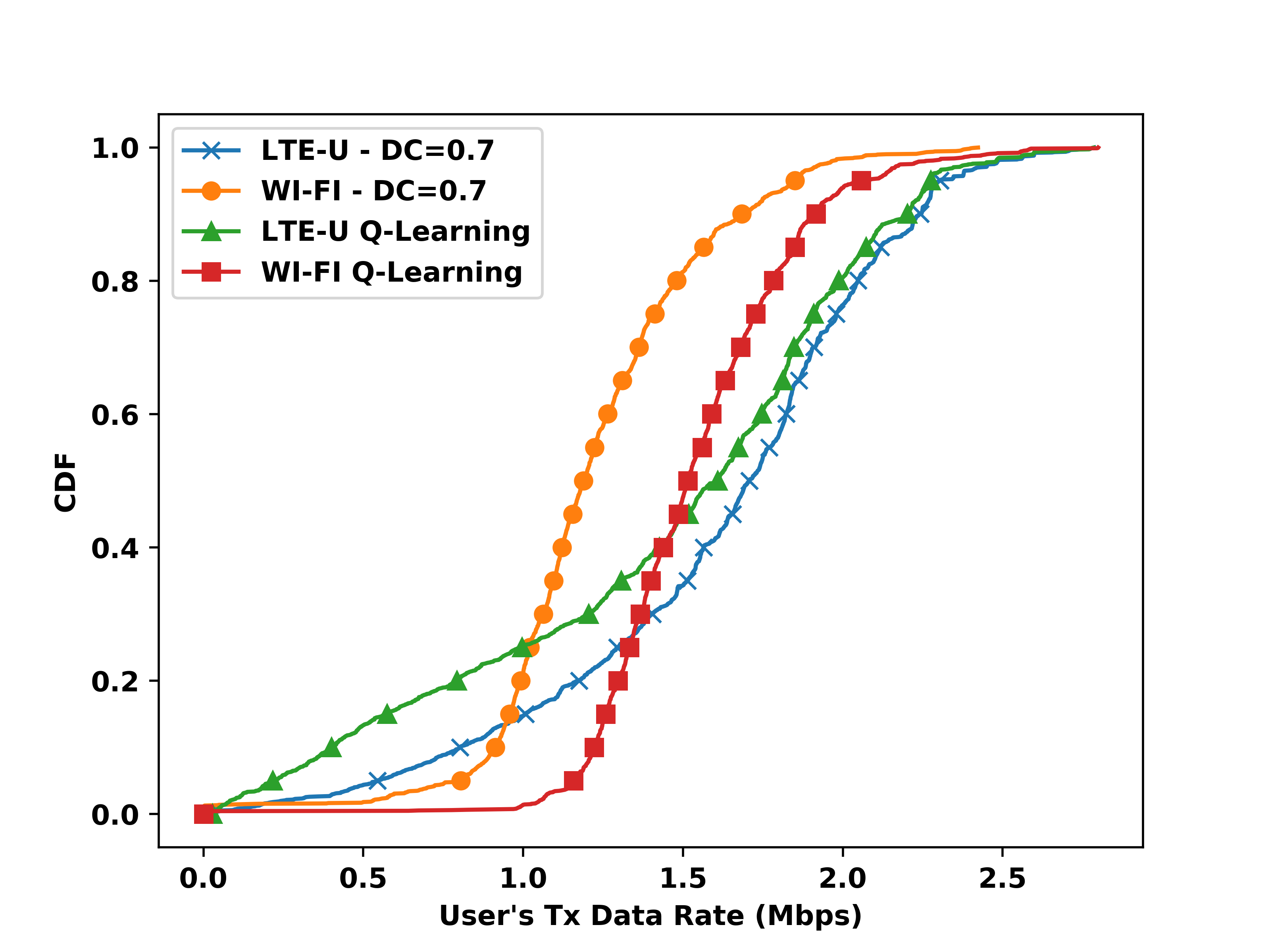}
  \caption{CDF of transmitted data rate per user vs Q-Learning for $DC=0.7$.}
  \label{fig:cdf_q_learning_0.7}
\end{figure}

For the duty cycle value of 0.7, the CDF presents a different behavior, as shown in Figure~\ref{fig:cdf_q_learning_0.7}. The Wi-Fi stations present even more gains with Q-Learning, despite the gain is comes with the cost of worse performance for the LTE-U stations. With Q-Learning, the 60th-percentile shows that when compared with the case of $DC=0.7$, the Wi-Fi gains affect the LTE-U stations with fewer data rates. Furthermore, the gain of $3$~Mbps presented at the system level, as opposed to the $DC=0.6$ case, is still distributed for practically all Wi-Fi stations. Still, the LTE-U stations with fewer data rates are severely~affected.  

%\clearpage

\begin{figure}[H]
  \centering
  \includegraphics[width=0.7\linewidth]{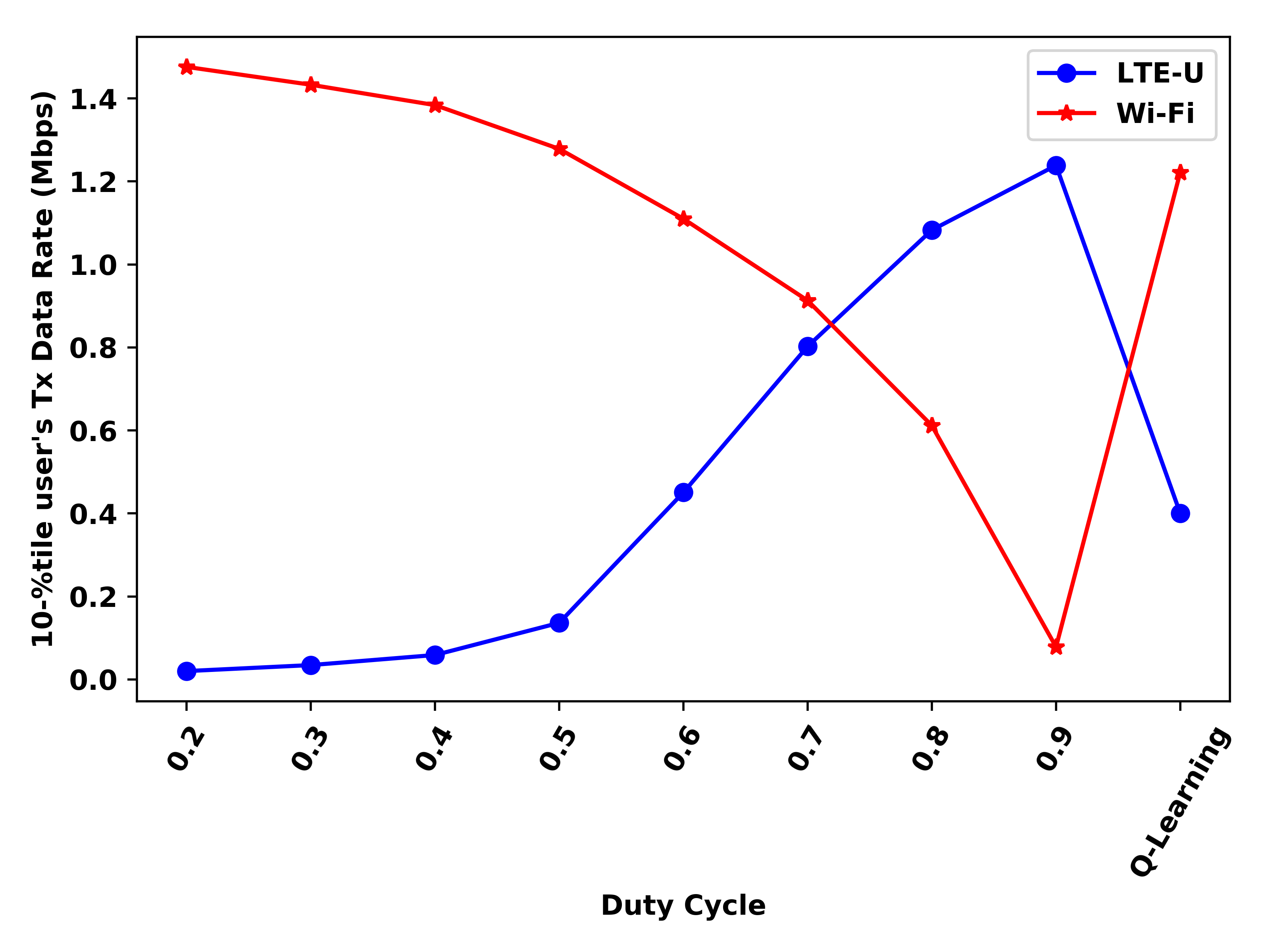}
  \caption{10th-percentile user's transmitted data rate.}
  \label{fig:10p_q_learning}
\end{figure}

\begin{figure}[H]
  \centering
  \includegraphics[width=0.7\linewidth]{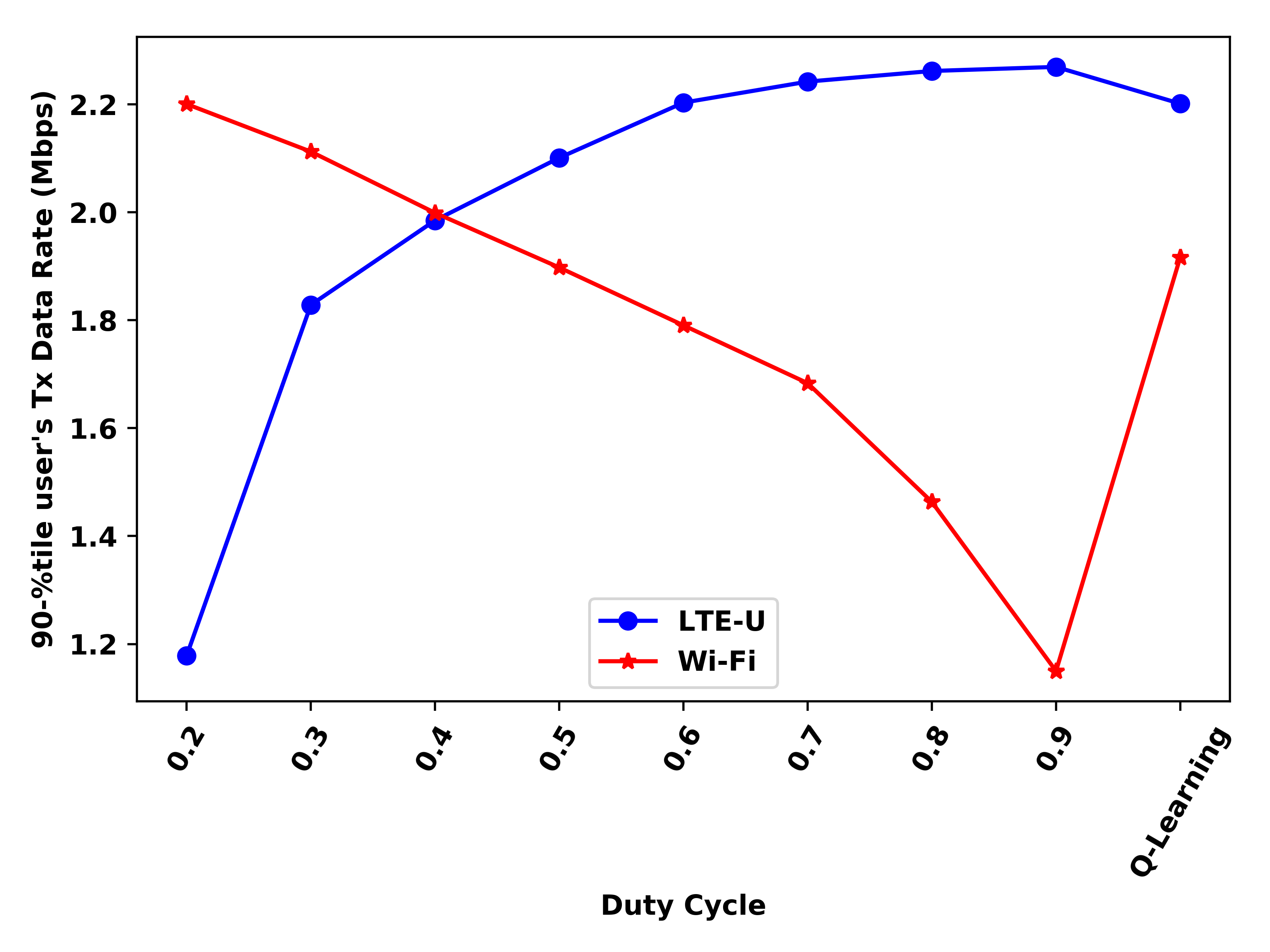}
  \caption{90th-percentile user's transmitted data rate.}
  \label{fig:90p_q_learning}
\end{figure}

Figure~\ref{fig:10p_q_learning}~and~\ref{fig:90p_q_learning} presents the comparison of the 10th and 90th percentile for each duty cycle value and the Q-Learning, respectively.

\clearpage

For the 10th-percentile, the Wi-Fi Q-Learning results are better than all duty cycles values higher than 0.5, while the LTE-U ones are better for the duty cycles values lower than 0.5.  For the 90th percentile, the Wi-Fi results with the proposed solution are better than all DC values higher than 0.4, while the LTE-U results are better for the duty cycle values lower than 0.5. Still, in the 90th-percentile analysis, it can be seen that LTE-U values do not show much variation for high values of the duty cycle. These duty cycles does not significantly affect this percentile for LTE-U, but only the Wi-Fi. 

\section{Conclusions and Future Works}
\label{conc}

In this work, we propose and evaluate a centralized, coordinated reinforcement learning framework, based on the Q-Learning algorithm, to deal with a multi-cell coexistence scenario between LTE-U and Wi-Fi. This scenario, differently from other previous works, is more challenging since it comprises not only more cells, but also interference between the same and different access technologies. Investigative results showed that each duty cycle value leads the system to a different maximum data rate, which also depends on the offered data rate we provided for each user in the coexistence scenario. 

The automatic duty cycle selection of our proposed solution is accomplished at a time window of 40~ms, reaching to a improved aggregated transmitted data rate. This gain is distributed all over the Wi-Fi users, with Q-Learning providing gains in both system and user transmitted data rate. % when Q-Learning is used, while the LTE-U users maintained theirs. These results also showed that even setting the goal to maximize the maximum aggregated transmitted data rate, the user's data rate is also affected in the right way.  

Regarding the multimedia IoT deployments, since our solution adopts a centralized and coordinated approach, the issues related to battery consumption are diminished because the additional energy consumption due to the Q-Learning algorithm is only applied to the central node that runs it. Furthermore, the benefits of our framework matches one of the goals of cellular broadband IoT deployments that is the use of data rates sufficient to transmit multimedia data while keeping a reasonable energy consumption.  Then, in the trade-off between energy consumption and increasing data rate, our solution achieves improvements in data rate without cost of energy consumption for each terminal. 

One drawback of our solution is that the way it was modeled, the interference between the same access technologies is not explicitly taken in account, even it is included in the $T_{tx \ wifi}$ and $T_{tx \ lteu}$ collected values due to the lost packets. If we also explicitly consider the interference measurement, the results could present even better results, since the algorithm would use this information to better perform in the choice of the duty cycle value. Then, we address this as an extension of our solution and a future work, when we also could explore modifications regarding QoS or a decentralized approach with minimum additional energy consumption.

\vspace{6pt} 

%%%%%%%%%%%%%%%%%%%%%%%%%%%%%%%%%%%%%%%%%%
%% optional
%\supplementary{The following are available online at \linksupplementary{s1}, Figure S1: title, Table S1: title, Video S1: title.}

% Only for the journal Methods and Protocols:
% If you wish to submit a video article, please do so with any other supplementary material.
% \supplementary{The following are available at \linksupplementary{s1}, Figure S1: title, Table S1: title, Video S1: title. A supporting video article is available at doi: link.}

%%%%%%%%%%%%%%%%%%%%%%%%%%%%%%%%%%%%%%%%%%
\authorcontributions{investigation, Jose Martins de Castro Neto and Sildolfo Francisco Gomes Neto; methodology, \hl{Vicente Angelo de Sousa Junior} and Jose Martins de Castro Neto; project administration, Vicente Angelo \hl{de} Sousa Junior; resources, Vicente Angelo De Sousa Junior; supervision, Pedro Maia de Santana and Vicente Angelo De Sousa Junior; validation, Vicente Angelo \hl{de} Sousa Junior; writing---original draft, Jose Martins de Castro Neto and Sildolfo Francisco Gomes Neto; Writing---review and editing, Jose Martins de Castro Neto, Sildolfo Francisco Gomes Neto, Pedro Maia de Santana and Vicente Angelo \hl{de} Sousa Junior.}

%%%%%%%%%%%%%%%%%%%%%%%%%%%%%%%%%%%%%%%%%%
\funding{This \hl{study} was \hl{financed} in part by the Coordena\c{c}\~{a}o de Aperfei\c{c}oamento de Pessoal de N\'{i}vel Superior - Brasil (CAPES) - Finance Code 001.
}
% Please add: ``This research received no external funding'' or ``This research was funded by NAME OF FUNDER grant number XXX.'' and  and ``The APC was funded by XXX''. Check carefully that the details given are accurate and use the standard spelling of funding agency names at \url{https://search.crossref.org/funding}, any errors may affect your future funding.}

%%%%%%%%%%%%%%%%%%%%%%%%%%%%%%%%%%%%%%%%%%
\acknowledgments{The proof of concept simulations provided by this paper was supported by the High-Performance Computing Center at UFRN (NPAD/UFRN).}

%%%%%%%%%%%%%%%%%%%%%%%%%%%%%%%%%%%%%%%%%%
\conflictsofinterest{The authors declare no conflict of interest.} 

\reftitle{References}

%%%%%%%%%%%%%%%%%%%%%%%%%%%%%%%%%%%%%%%%%%
%% optional
%\sampleavailability{Samples of the compounds ...... are available from the authors.}

%% for journal Sci
%\reviewreports{\\
%Reviewer 1 comments and authors’ response\\
%Reviewer 2 comments and authors’ response\\
%Reviewer 3 comments and authors’ response
%}

%%%%%%%%%%%%%%%%%%%%%%%%%%%%%%%%%%%%%%%%%%
\end{document}